%% file: main.tex
\documentclass[conference]{llncs}

\usepackage{amsmath,amssymb,amsfonts}
\usepackage{algorithmic}
\usepackage{graphicx}
\usepackage{textcomp}
\usepackage{xcolor}
\usepackage{subcaption}

\definecolor{oliv}{HTML}{A491D3}
\definecolor{lightgold}{HTML}{EAC71B}
\definecolor{sandybrown}{HTML}{D76C0E}
\definecolor{jungleteal}{HTML}{697A21}
\definecolor{blushedbrick}{HTML}{863136}

\usepackage{orcidlink} 
\usepackage[T1]{fontenc}
\usepackage{paralist} 
\usepackage{makecell}
\usepackage{booktabs}

\usepackage{float}

\usepackage{pdflscape}   
\usepackage{longtable}
\usepackage{multirow}
\usepackage{tikz}
\usetikzlibrary{calc,positioning,arrows.meta} 

\usepackage{pgfplots}
\pgfplotsset{compat=1.18}
\usepgfplotslibrary{groupplots}

\newcolumntype{P}[1]{>{\centering\arraybackslash}m{#1}}
\newcolumntype{L}[1]{>{\raggedright\arraybackslash}m{#1}}

\begin{document}

\title{Few-Shot Learning for Network Intrusion Detection: Methods, Datasets, and Performance}

\author{Arne Roszeitis \and Victor Jüttner \and Erik Buchmann}
\institute{Center for Scalable Data Analytics and Artificial Intelligence (ScaDS.AI) Dresden/Leipzig, Leipzig University, Leipzig, Germany \\
Email: \{arne.roszeitis, victor.juettner, erik.buchmann\}@uni-leipzig.de}

\maketitle
\input{00-abstract}
\input{01-introduction}
\input{02-related}

\input{03-method-new}
\input{04a-general-results}
\input{04b-learning-techniques}
\input{05-discussion-new}

\input{06-conclusion}
\input{07-acknowledgements} 


\bibliographystyle{splncs04}
\bibliography{literature}

\input{08-appendix}

\end{document}

%% file: 00-abstract.tex
\begin{abstract}

Anomaly-based network intrusion detection systems (NIDS) are an important first line of defense. 
However, training NIDS for new attack types is challenging, because labeled attack data are rarely available. Few-shot learning (FSL) addresses this problem by learning from few samples. However, the approaches and evaluation settings, that have been investigated so far, vary widely. 
This work systematically reviews FSL approaches for NIDS published from 2022 to 2026. 
We conduct a systematic literature review with PRISMA 2020-like reporting to search ACM Digital Library, IEEE Xplore, and Scopus. From a set of 1,358 initial records, we retain 21 studies after screening, deduplication, and quality filtering. We classify the applied FSL approaches, datasets, and experimental parameters and compare reported performance. Meta-learning and convolutional neural networks are the most common approaches, with 8 and 10 studies, respectively. Most studies evaluate five or fewer samples per class, although settings vary. CIC-IDS2017 and CSE-CIC-IDS2018 are the most frequently used datasets. Missing parameters and source code limit reproducibility and direct comparison between approaches.
\end{abstract}

\keywords{Network Intrusion Detection  \and Few-Shot Learning \and Review}

%% file: 01-introduction.tex
\section{Introduction}
\label{sec:intro}

Network intrusion detection systems~\cite{Khraisat2019SurveyOIA} (NIDS) identify attacks that cross network boundaries, complement host-based defenses, and provide control points in heterogeneous environments~\cite{Ahmed2016ASOA}. As today's attackers increasingly use generative AI~\cite{gupta2023chatgpt} to modify payloads, alter protocol interactions, and tailor attacks to specific systems~\cite{Ahmed2025NetworkIDA}, NIDS must adapt quickly to new attacks.

Signature-based detection cannot recognize attacks for which no rule exists, and anomaly-based approaches raise alerts without identifying the attack type~\cite{Khraisat2019SurveyOIA,Ahmed2016ASOA}.
Supervised, learning-based NIDS close this gap by generalizing from labeled traffic; however, they require labeled samples of every attack class they are expected to recognize.
For novel attacks, such labels are scarce, while benign traffic dominates the available data~\cite{Garg2026AROA}. Few-shot learning~\cite{wang2020generalizing} (FSL) addresses exactly this situation by learning new classes from a small number of labeled samples, making it a natural fit for intrusion detection under label scarcity.

Driven by this promise, a growing body of studies has applied FSL to network intrusion detection in recent years.
However, the field has not yet converged on common practices. There is no consensus on which FSL techniques or combinations perform best, nor on how they should be evaluated. Moreover, the meaning of ``few'' remains inconsistent, with recent studies using between one and 20 samples per class~\cite{ye2022dirichlet,zhang2025transferlearning}. Studies differ in the number of samples per class, the number of classes, the datasets used, preprocessing steps, and reported metrics, and reporting practices vary widely. Without such common ground, reported results cannot be compared across studies and reproduction is difficult wherever parameters remain undisclosed.
To structure the current landscape, we answer three research questions:
\begin{description}
\item[RQ1] \textit{Which learning techniques are used for few-shot NIDS?}
\item[RQ2] \textit{Which datasets are used to evaluate these approaches?}
\item[RQ3] \textit{How are these approaches evaluated?} 
\end{description}

In this paper, we contribute a systematic literature review following~\cite{kitchenham2007guidelines} with a PRISMA 2020-like reporting~\cite{page2021prisma} of recent studies on FSL for NIDS. We searched the ACM Digital Library, IEEE Xplore, and Scopus for peer-reviewed studies published between 2022 and 2026. Our search returned 1,358 records, of which we retained 21 studies for detailed analysis after deduplication and filtering according to predefined  criteria. For each study, we extract the applied learning technique, datasets, evaluation settings, and reported FSL parameters where available. We organize the studies by their main techniques and compare their reported results where possible.

Most reviewed studies combine multiple learning techniques (17 of 21), with CNNs (10 of 21) and meta-learning (8 of 21) being the most common. Among the 17 studies reporting \(k\), \(k{=}5\) is the most frequent setting. CIC-IDS2017 and CSE-CIC-IDS2018 are the most widely used datasets, appearing in 11 and 6 studies, respectively. Accuracy, precision, recall, and F1-score are the dominant evaluation metrics, with 15 studies reporting F1-scores. However, evaluation settings vary substantially: five studies omit \(k\), four omit the number of classes \(n\), and most provide no accessible source code. These gaps limit reproducibility and direct comparison.

%% file: 02-related.tex
\section{Related Work}




\paragraph{Intrusion detection} has been surveyed extensively. Aldweesh et al.~\cite{aldweesh2020surveyDL} 
provide a taxonomy of deep-learning approaches for anomaly-based IDS together with open issues. 
Maseer et al.~\cite{maseer2024metaReviewIDS} present a meta review of NIDS, finding deep learning approaches to perform better than traditional machine learning. Also, they criticize the use of out-of-date data sets.
More recently, Lansky et al.~\cite{lansky2021reviewDL} provide an in-depth review of IDS systems and their deep learning architectures. 
Ashraf and Masoodi ~\cite{ashraf2026surveyIDSdatasets} 
examine over 120 data sets used for intrusion detection with regard to attack types, temporal features and other categories. 
Few-shot learning was only once mentioned in \cite{aldweesh2020surveyDL} and \cite{lansky2021reviewDL}; and not at all in \cite{maseer2024metaReviewIDS}.  
\paragraph{Few-Shot Learning Techniques}
Coming from the other angle, few-shot learning itself has been systematized by Wang et al.~\cite{wang2020generalizing}, who categorize FSL techniques by how prior knowledge is used to augment data, constrain the model, or guide the optimization algorithm. 
Song et al.~\cite{song2023comprehensive} review over 200 recent FSL studies and demarcate few-shot learning from the related concepts of transfer learning and meta-learning. 
For the related problem of continuously arriving new classes, Zhou et al.~\cite{zhou2024class} survey class-incremental learning. For the neighboring task of encrypted traffic classification, Yang et al.~\cite{yang2024encryptedSurvey} survey few-shot approaches specifically. 
These studies are not primarily concerned with network intrusion, however, Yang et al.~\cite{yang2024encryptedSurvey} are very close. 

\paragraph{Closest Work}Even closer to our scope, Duan et al.~\cite{duan2021fslids} review few-shot learning for intrusion detection, covering data enrichment, graph embedding, and meta-learning approaches published up to 2021. 
Independent of our systematic search, Winiecki et al.~\cite{winiecki2025survey} evaluate selected FSL techniques, including a matching network with attention and prototyping, against classical baselines such as kNN and random forests for network intrusion detection.
\paragraph{Research Gap} In summary, existing surveys either address intrusion detection broadly without a few-shot focus,  or, in the case of Duan et al.~\cite{duan2021fslids}, predate the rapid development of the field after 2021. To the best of our knowledge, no systematic review of few-shot learning for network intrusion detection exists since then. This work closes that gap by systematically reviewing studies published between 2022 and 2026, with a focus on the applied techniques, the datasets, and the evaluation settings, which together determine how comparable the reported results are.

%% file: 03-method-new.tex
\section{Methodology}
\label{sec:method}

We conduct a systematic literature review following~\cite{kitchenham2007guidelines}, with a PRISMA 2020-like reporting~\cite{page2021prisma}. 
Our study selection process is illustrated in Figure~\ref{fig:prisma_decision_process}. 
We queried the ACM Digital Library~\cite{acm}, IEEE Xplore~\cite{ieee}, and Scopus~\cite{scopus}.  
Each query combines intrusion detection and prevention terms (\textit{intrusion detection}, \textit{IDS}, \textit{intrusion prevention}, \textit{IDPS}) with few-shot and one-shot learning terms (\textit{few shot learning}, \textit{FSL}, \textit{one-shot learning}, \textit{one shot learning}, \textit{OSL}), because OSL is contained in FSL.  
The broader term \textit{learning} was included to capture studies not using these terms in the indexed metadata. 
The full queries are shown in the Appendix.

\input{fig-paperselection}

The search returned 1034 (ACM), 183 (IEEE), and 141 (Scopus) records. 
We compared these 1358 studies against the eligibility criteria listed in Table~\ref{tab:eligibility_criteria}. 
In particular, we automatically filtered titles and abstracts for the terms \emph{intrusion}, \emph{detection}, \emph{shot}, and \emph{learning}, all of which had to match, leaving 162 studies. After deduplication, 124 studies remained. Of these, 34 were published in a venue meeting our quality criterion, i.e., a journal with an $h_4$-index of 53 or above as listed in the OOIR database~\cite{ooir} or a conference ranked B or better in the CORE ranking~\cite{icore}. The $h_4$-index denotes the maximum number of a journal's studies with at least that many citations within the last four years.

\begin{table}[H]
\centering
\caption{Eligibility criteria for included studies.}
\label{tab:eligibility_criteria}
\small
\setlength{\tabcolsep}{5pt}
\begin{tabular}{@{}p{2.7cm}p{8.5cm}@{}}
\toprule
\textbf{Criterion} & \textbf{Requirement} \\
\midrule
Publication period
    & January 2022 -- August 2026 \\

Language \& Type
    & English peer-reviewed research\\

Study
    &  Few-shot or one-shot learning applied to NIDS \\

Keywords
    &  \emph{intrusion}, \emph{detection}, \emph{shot}, and \emph{learning} \\

Venue
    & Journal $h_4$-index $\geq 53$ or CORE-ranked conference $\geq$ B \\

Datasets
    & Excluded if  $\geq 50\%$ of evaluated datasets predate 2007 \\
\bottomrule
\end{tabular}
\end{table}

After automated filtering, we manually tested the remaining 34 studies for eligibility against the criteria listed in Table~\ref{tab:eligibility_criteria}, resulting in 21 studies being included in the review. Four of the excluded studies evaluated  datasets predating 2007, six were not concerned with few-shot learning for network intrusion detection or reported no information about it, one targeted zero-day attacks, which is out of scope, and one did not evaluate its approach conclusively. The remaining exclusion was a review, not a study; while it presents no novel approach itself, we reference it for context~\cite{chen2026gnn}.

For each included study, we extracted the applied learning techniques, the datasets used for evaluation, the evaluation setting (the number of classes $n$ and samples per class $k$, where reported), and the reported performance. 
Most studies evaluate their approach on more than one dataset and report different subsets of metrics.
For studies evaluating multiple settings, we report the performance of the 5-shot setting ($k{=}5$), as this is the most commonly evaluated setting. If no 5-shot experiment is reported, we use the study's best-performing setting.
All numbers are truncated after the third decimal place. The full per-dataset results at the selected setting, including accuracy, precision, recall, and F1-score, are given in the Appendix. 

We structure the review by grouping studies according to their main FSL techniques. The set of techniques was derived from the corpus itself. 

\input{tab-papersincluded}

In a first pass over all included studies, we listed every learning technique applied, and then iteratively merged 
synonymous and subordinate techniques until ten categories remained that together cover every but one included studies. 
The categories are neither exclusive nor exhaustive. Most studies combine several FSL techniques. Table~\ref{tab:topics} lists the techniques per study and Figure~\ref{fig:technique-trends-and-topics} shows their co-occurrence and usage per year.

%% file: fig-paperselection.tex
\begin{figure}[H]
\centering
\resizebox{\linewidth}{!}{%
\begin{tikzpicture}[
    font=\sffamily\small,
    node distance=4mm and 6mm,
    box/.style={
        draw=black!60, line width=0.6pt, rounded corners=3pt,
        fill=black!4, align=center, text width=2.3cm,
        minimum height=2.1cm, inner sep=4pt,
        anchor=center
    },
    exbox/.style={
        draw=black!45, line width=0.5pt, rounded corners=3pt,
        fill=black!2, align=center, text width=2.1cm,
        minimum height=0.85cm, inner sep=3pt, font=\sffamily\small
    },
    tag/.style={
        draw=none, fill=black!75, text=white, rounded corners=5pt,
        inner sep=3pt, font=\sffamily\small\bfseries, align=center, anchor=south,
        text height=1.5ex, text depth=0.25ex
    },
    arr/.style={-{Triangle[length=2mm,width=1.4mm]}, line width=0.6pt, draw=black!70},
    exarr/.style={-{Triangle[length=1.6mm,width=1.2mm]}, line width=0.5pt, draw=black!45}
]

\node[box] (n1) {Database search\\[2pt]\textbf{(No=1358)}\\[2pt] ACM:1034, IEEE:183,\\Scopus:141};
\node[box, right=of n1] (n2) {Keyword search \&\\publication criteria\\[2pt]\textbf{(No=162)}};
\node[box, right=of n2] (n3) {After\\deduplication\\[2pt]\textbf{(No=124)}};
\node[box, right=of n3] (n4) {Full-text\\eligibility check\\[2pt]\textbf{(No=34)}};
\node[box, right=of n4] (n5) {Full-text articles\\included\\[2pt]\textbf{(No=21)}};

\draw[arr] (n1) -- (n2);
\draw[arr] (n2) -- (n3);
\draw[arr] (n3) -- (n4);
\draw[arr] (n4) -- (n5);

\node[tag] at ([yshift=3mm]n1.north) {Identification};
\node[tag] at (n2 |- n1.north) [yshift=3mm] {Screening};
\node[tag] at (n3 |- n1.north) [yshift=3mm] {Deduplication};
\node[tag] at (n4 |- n1.north) [yshift=3mm] {Eligibility};
\node[tag] at (n5 |- n1.north) [yshift=3mm] {Inclusion};

\node[exbox, below=6mm of n2] (e1) {Excluded\\\textbf{(No=1053)}};
\node[exbox, below=6mm of n3] (e2) {Excluded\\\textbf{(No=90)}};
\node[exbox, below=6mm of n4] (e3) {Excluded\\\textbf{(No=13)}};

\draw[exarr] (n2.south) -- (e1.north);
\draw[exarr] (n3.south) -- (e2.north);
\draw[exarr] (n4.south) -- (e3.north);

\end{tikzpicture}%
}
\caption{Our study selection process}
\label{fig:prisma_decision_process}
\end{figure}
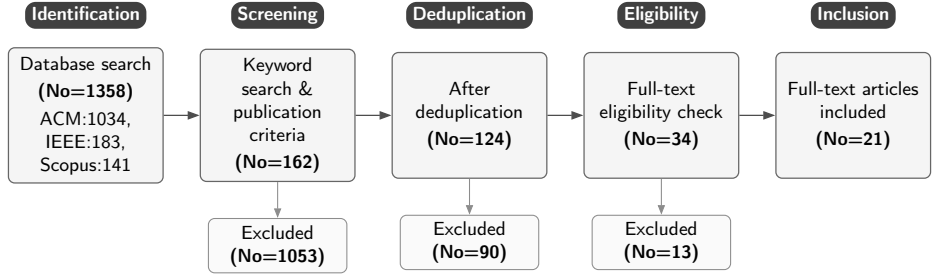

%% file: tab-papersincluded.tex
{
\small

\begin{longtable}{P{1.25cm}P{3.7cm}P{2.25cm}cccccccccc}
\caption{Publication year, number of datasets and learning techniques.}
\label{tab:topics} \\

\toprule
Year & Study & No. Datasets &
\rotatebox{90}{Meta-Learning} &
\rotatebox{90}{MAML} &
\rotatebox{90}{CNN} &
\rotatebox{90}{Auto-Encoder} &
\rotatebox{90}{FSCIL} &
\rotatebox{90}{Twin NN} &
\rotatebox{90}{Self-supervised Learning} &
\rotatebox{90}{Data generation} &
\rotatebox{90}{Attention} &
\rotatebox{90}{Graph NN} \\
\midrule
\endfirsthead

\multicolumn{13}{c}%
{{\tablename\ \thetable{} -- continued from previous page}} \\
\toprule
Year & Study & No. Datasets &
\rotatebox{90}{Meta-Learning} &
\rotatebox{90}{MAML} &
\rotatebox{90}{CNN} &
\rotatebox{90}{Auto-Encoder} &
\rotatebox{90}{FSCIL} &
\rotatebox{90}{Twin NN} &
\rotatebox{90}{Self-supervised Learning} &
\rotatebox{90}{Data generation} &
\rotatebox{90}{Attention} &
\rotatebox{90}{Graph NN} \\
\midrule
\endhead

\midrule
\multicolumn{13}{r}{{Continued on next page}} \\
\endfoot

\bottomrule
\endlastfoot

2026 & Martinez-Lopez et al. \cite{martinez2026multipleSpaces} & 4
& & & & & & & & \checkmark & & \\

& Wu et al. \cite{wu2026negativeLearning} & 3
& \checkmark & \checkmark & \checkmark & & & & & & & \\

& Yin et al. \cite{yin2026efficient} & 2
& & & \checkmark & & \checkmark & & & & & \\

& Jamshidi et al. \cite{jamshidi2026thinkFast} & 1
& & & & & & & & & & \\

& Zhang et al. \cite{zhang2026maml-samIoT} & 2
& \checkmark & \checkmark & & & & & & & & \\

& Asante and Abass \cite{asante2026robustMalwareDetection} & 1
& \checkmark & & & & & & & & & \checkmark \\

& Lu et al. \cite{lu2026dmaml} & 3
& \checkmark & \checkmark & & & & & & & & \\

\midrule

2025 & Mao et al. \cite{mao2025federated} & 3
& & & & & & & \checkmark & & & \checkmark \\

& Zhang et al. \cite{zhang2025transferlearning} & 3
& & & \checkmark & & & \checkmark & & & & \\

& Xu et al. \cite{xu2025multimodalFusion} & 2
& & & \checkmark & & & & & \checkmark & \checkmark & \\

& Xu et al. \cite{xu2025mutalCentralizedLearning} & 3
& \checkmark & & \checkmark & & & & & \checkmark & & \\

& Qiu et al. \cite{qiu2025disentaglement} & 5
& & & & & & & & & & \checkmark \\

\midrule

2024 & Tong and Zhang \cite{tong2024realTimeLabelFree} & 3
& & & \checkmark & \checkmark & & & \checkmark & & \checkmark & \\

& Du et al. \cite{du2024fscilForNIDS} & 2
& & & & \checkmark & \checkmark & & \checkmark & & \checkmark & \\

\midrule

2023 & Ayesha et al. \cite{ayesha2023fs3} & 3
& & & & \checkmark & & & \checkmark & & & \\

& Hang et al. \cite{hang2023FlowMAE} & 5
& & & & \checkmark & & & \checkmark & & \checkmark & \\

& Lu et al. \cite{lu2023MAML_IDS} & 1
& \checkmark & \checkmark & \checkmark & & & & & & & \\

& Mirsadeghi et al. \cite{mirsadeghi2023sdn} & 1
& & & \checkmark & & & \checkmark & & \checkmark & & \\

& Miao et al. \cite{miao2023spn} & 3
& \checkmark & & \checkmark & & & \checkmark & & & & \\

& Sun et al. \cite{sun2023capsuleNetwork} & 1
& \checkmark & & \checkmark & & & & & & \checkmark & \\

\midrule

2022 & Ye et al. \cite{ye2022dirichlet} & 2
& & & & & & & & \checkmark & & \\

\midrule

Count & 21 & & 8 & 4 & 10 & 4 & 2 & 3 & 5 & 5 & 5 & 3 \\

\end{longtable}

}

%% file: 04a-general-results.tex
\section{Review Results}
\label{sec:results}

This section provides an overview of our review, lists the datasets and evaluation settings used, and then examines the FSL technique families in more detail.

\subsection{Overview}

Table~\ref{tab:topics} is the central artifact of this review: It provides an overview of the 21 included studies, along with the publication date, the learning techniques combined in each and the number of datasets evaluated. Full per-dataset results, including the evaluation parameters $n$ and $k$, are provided in the Appendix.

\begin{figure}[h]
  \centering
  \includegraphics[width=\textwidth]{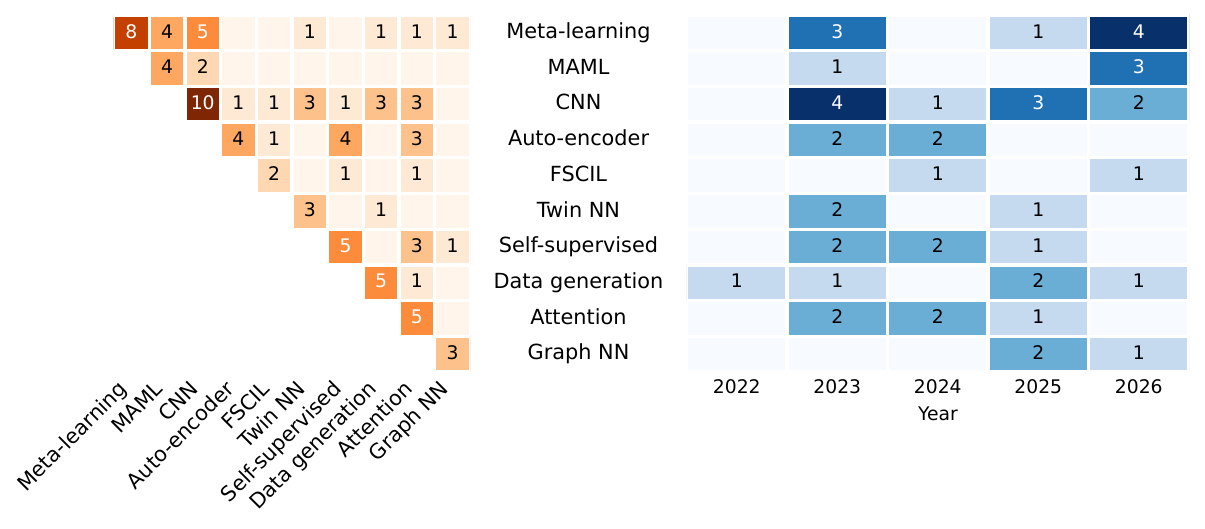}
  \caption{Left: Topic co-occurrences. Diagonal is the total number of occurrences of this topic. Right: Technique usage per year.}
  \label{fig:technique-trends-and-topics}
\end{figure}

Figure~\ref{fig:technique-trends-and-topics} summarizes the combinations of learning techniques and their temporal distribution. The left panel shows pairwise co-occurrences between techniques; diagonal entries indicate the total number of studies using each technique, while off-diagonal entries show how often two techniques are combined. CNNs (10 studies) and meta-learning (8 studies) are the most common, and their combination occurs most frequently (5 studies). Overall, 17 of the 21 studies combine at least two techniques. The right panel shows the number of studies using each technique per publication year as a heatmap, with darker cells indicating higher frequencies. CNNs are used throughout the reviewed period, while meta-learning is most frequent in 2026 and graph neural networks appear only in 2025--2026.

Figure~\ref{fig:technique-trends-and-topics} further shows that approaches for learning from few labeled samples (Meta-learning, MAML, FSCIL, and Twin NN) are more common than data-generation approaches for augmenting limited training data. 
The temporal distribution also shows a shift in focus. Meta-learning is absent in 2024 and appears only once in 2025 before increasing again in 2026. CNNs are used throughout 2023--2026, while Auto-Encoders appear only in 2023 and 2024 despite the strong performance reported by the respective studies.

\subsection{Datasets}
To characterize the empirical basis of the reviewed studies, we next examine the datasets used for evaluation. We consider both their frequency of use and their overlap across studies, as common datasets provide the main basis for comparing reported results.
Table~\ref{tab:dataset_frequency} summarizes how often which dataset was used in a study; the appendix contains all details. 

\input{tab-datasetsused}

CIC-IDS2017 \cite{CIC-IDS2017} and CSE-CIC-IDS2018 \cite{CSE-CIC-IDS2018} were by far the most popular choices (11 and 6 uses), followed by USTC-TFC2016 \cite{USTC-TFC2016} and UNSW-NB15 \cite{UNSW-NB15} (5 and 1 uses). Additionally, CIC-IDS2017, CSE-CIC-IDS2018 and UNSW-NB15 were used in other datasets (FSIDS-IoT(-v2), prefixed NF-[dataset](-v2)), adding between two and five uses depending on the counting method. Each study evaluated a median of three datasets. At the same time, the long tail of single-use datasets  means that many results cannot be compared against any other study.

\subsection{Evaluation and Performance}
We next examine the evaluation settings reported by the included studies. Figure~\ref{fig:nk-overview} summarizes the $n$-way and $k$-shot configurations. Only 16 of the 21 studies explicitly specify $k$. Of these, 12 evaluate settings with five or fewer samples per class. The most common setting is $k{=}5$. The number of classes is typically between five and eight, with values ranging from two to 18. 
Five studies do not report $k$ and five do not report $n$, limiting reproducibility and comparability.

\input{fig-nk}

Regarding reported performance, 15 of the 21 studies provide an F1-score. Their best reported F1-scores range from 0.700 to 0.999, and 12 of these 15 studies achieve at least 0.900 on one dataset. Five studies~\cite{ye2022dirichlet,xu2025multimodalFusion,qiu2025disentaglement,yin2026efficient,asante2026robustMalwareDetection} report particularly high performance $F_1 \approx 0.96$, $Acc \approx 0.96$, $F_1 \approx 0.97$, $Precision \approx 0.98$, $Precision = 0.996$). 

However, these results are difficult to compare directly, even when using the same dataset. Among the six studies reporting an F1-score on CIC-IDS2017, values range from 0.673 to 0.995. Similar variation appears within individual studies: Ayesha et al.~\cite{ayesha2023fs3} report F1-scores between 0.596 and 0.980 across three datasets, while Tong and Zhang~\cite{tong2024realTimeLabelFree} report 0.913 on NSL-KDD and 0.997 on CSE-CIC-IDS2018 without specifying $k$. These differences indicate that dataset choice and evaluation settings substantially affect the reported performance and limit comparisons between techniques.

%% file: tab-datasetsused.tex
\begin{table}[htbp]
\centering
\footnotesize
\caption{Frequency of the datasets used.}
\label{tab:dataset_frequency}
\begin{tabular}{p{9cm}c}
\toprule
\textbf{Dataset} & \textbf{Freq.} \\
\midrule
CIC-IDS2017     & 11 \\ \midrule
CSE-CIC-IDS2018 & 6  \\ \midrule
USTC-TFC2016    & 5  \\ \midrule
Edge-IIoT, CIC-BoT-IoT       & 3  \\ \midrule
FSIDS-IoT, IoT-23, ISCX-IDS2012,
NF-CSE-CIC-IDS2018-v2, CIC-ToN-IoT
& 2 \\ \midrule
CIC-EVSE2024 Network, CIC-EVSE2024 PowerB,
CIC-IoV2024, Cross-platform, FSIDS-IoT-v2,
InSDN, ISCX-Tor-2016, ISCX-VPN-2016 (App),
ISCX-VPN-2016 (Service), NF-BoT-IoT-v2, NF-ToN-IoT-v2,
NF-UNSW-NB15-v2, NSL-KDD, UNSW-NB15,
WuSTl-EHMS, WuSTl-IIoT
& 1 \\ 
\bottomrule
\end{tabular}
\end{table}


%% file: fig-nk.tex
\definecolor{accent}{HTML}{6D4AFF}
\begin{figure}
\centering
    \caption{Reported evaluation parameters per study. Left: number of classes $n$; right: samples per class $k$. Studies not reporting a parameter are omitted.}
    \label{fig:nk-overview}
\begin{tikzpicture}
\begin{groupplot}[
    group style={
        group size=2 by 1,
        horizontal sep=0.6cm,
    },
    height=8cm,
    scale only axis,
    width=4cm,
    xmin=0, xmax=22,
    grid=major,
    major grid style={gray!20},
    tick label style={font=\small},
    label style={font=\small},
    axis line style={black},
    every axis plot/.append style={
        only marks, mark=*, mark size=2.2pt, color=accent,
    },
]

\nextgroupplot[
    xlabel={n (classes)},
    ylabel style={font=\small},
    ytick={0,1,2,3,4,5,6,7,8,9,10,11,12,13,14,15,16,17,18,19,20},
    yticklabels={
        Lu~'26~\cite{lu2026dmaml},  {Asante and Abass~'26}~\cite{asante2026robustMalwareDetection},  Zhang~'26~\cite{zhang2026maml-samIoT},  Jamshidi~'26~\cite{jamshidi2026thinkFast},  Yin~'26~\cite{yin2026efficient}, Wu~'26~\cite{wu2026negativeLearning},  {Martinez-Lopez~'26}~\cite{martinez2026multipleSpaces},  Qiu~'25~\cite{qiu2025disentaglement},  Xu~'25~\cite{xu2025mutalCentralizedLearning},  Xu~'25~\cite{xu2025multimodalFusion}, Zhang~'25~\cite{zhang2025transferlearning},  Mao~'25~\cite{mao2025federated},  Du~'24~\cite{du2024fscilForNIDS},  {Tong and Zhang~'24}~\cite{tong2024realTimeLabelFree},  Sun~'23~\cite{sun2023capsuleNetwork}, Miao~'23~\cite{miao2023spn},  Mirsadeghi~'23~\cite{mirsadeghi2023sdn},  Lu~'23~\cite{lu2023MAML_IDS},  Hang~'23~\cite{hang2023FlowMAE},  Ayesha~'23~\cite{ayesha2023fs3}, Ye~'22~\cite{ye2022dirichlet}
    },
    y tick label style={font=\scriptsize, align=right},
]
\addplot coordinates {
    (2,0) (5,0) (5,1)  (5,2)  (15,3) (5,4)
    (5,5) 
    (2,8) (4,8) (5,8) (2,9)(3,9)(4,9)
    (5,10)(5,11) (7,11) (10,11) (12,12) (14,12) (5,13) (7,13) (8,14)
    (8,16) (5,17) 
    (2,19) (5,19) 
    (18,20)
};

\node[anchor=west, font=\tiny, text=red]
    at (axis cs:11,6) {no number given};
\node[anchor=west, font=\tiny, text=red]
    at (axis cs:11,7) {no number given};
\node[anchor=west, font=\tiny, text=red]
    at (axis cs:11,15) {no number given};
\node[anchor=west, font=\tiny, text=red]
    at (axis cs:11,18) {no number given};

\nextgroupplot[
    xlabel={k (samples per class)},
    ytick=\empty,                 
    clip=false,
]
\addplot coordinates {
    (5,0)  (1,1)(5,1) (1,2)(5,2)(10,2)  
    (5,4) 
    (5,5) (10,5) (2,6) (5,7) (5,8)(10,8) (5,9)(10,9)(15,9) (1,10)(5,10)(10,10)(20,10)
    (5,12) 
    (1,15)(3,15)(5,15)(10,15) (1,16) (1,17)(5,17)(10,17) 
    (5,19)(10,19) (1,20)(10,20)
};
\draw[dashed, red] (axis cs:5,\pgfkeysvalueof{/pgfplots/ymin}) -- (axis cs:5,\pgfkeysvalueof{/pgfplots/ymax});
\node[anchor=west, font=\small, text=red]
    at (axis cs:5.2,21) {k=5};
\node[anchor=west, font=\tiny, text=red]
    at (axis cs:11,3) {no number given};
\node[anchor=west, font=\tiny, text=red]
    at (axis cs:11,11) {no number given};
\node[anchor=west, font=\tiny, text=red]
    at (axis cs:11,13) {no number given};
\node[anchor=west, font=\tiny, text=red]
    at (axis cs:11,14) {no number given};
\node[anchor=west, font=\tiny, text=red]
    at (axis cs:11,18) {no number given};
\end{groupplot}
\end{tikzpicture}
\end{figure}
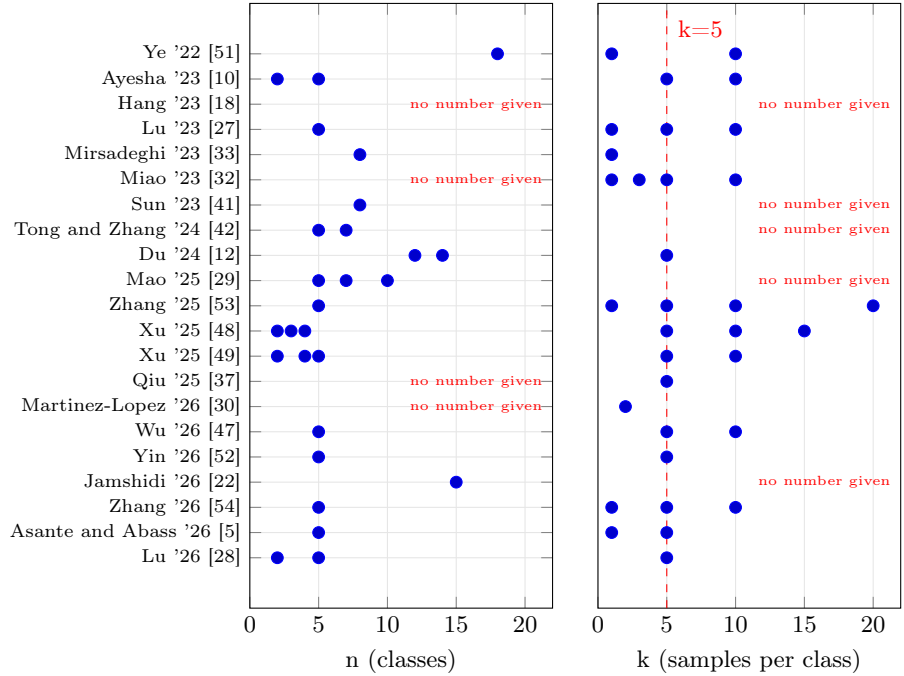

%% file: 04b-learning-techniques.tex
\subsection{Technique-Level Analysis}
\label{sec:majApproach}

This subsection examines the main learning techniques in detail. For each technique family, we introduce its application in few-shot intrusion detection, describe how it is used in the studies, and summarize the reported results.

\paragraph{Meta-Learning} trains an outer optimization loop over many small tasks so that the resulting model adapts to new classes from only a few labeled samples; the MAML framework \cite{finn2017maml} is the most common instantiation, used by four of the eight meta-learning studies. Inner learning rates of $\alpha = 0.01$ and meta-learning rates $\beta = 0.001$ where reported. Two MAML studies pair the framework with a CNN as the inner model. Lu et al.~\cite{lu2023MAML_IDS} convert flows into 3-channel images and optimize the CNN through MAML in a 5-way setting, while Zhang et al.~\cite{zhang2026maml-samIoT} move detection to fog devices and extend MAML with multi-step loss optimization and a learn-to-forget mechanism. Wu et al.~\cite{wu2026negativeLearning} pseudo-label formerly unlabeled data via negative learning and use bagging against class imbalance. Easing the few-shot problem itself, Lu et al.~\cite{lu2026dmaml} generate training samples with a denoising diffusion model \cite{ho2020ddpm} enhanced by a drift loss. The remaining meta-learning studies constructed other frameworks: Miao et al.~\cite{miao2023spn} train twin 3D-CNNs to find and handle out-of-distribution classes through prototype distances, Sun et al.~\cite{sun2023capsuleNetwork} combine capsule networks to extract spatial and temporal features, as well as attention-based prototypes, and Xu et al.~\cite{xu2025mutalCentralizedLearning} use bidirectional matching between labels and flows via cosine similarity and a random walk. The results within this family vary widely: where F1-scores are reported, they range from 0.915 \cite{zhang2026maml-samIoT} to 0.996 \cite{asante2026robustMalwareDetection}. Meanwhile, the MAML-using studies are located in a narrower band between 0.915 and 0.960. Meta-learning is elegant, but appears hard to get right in practice.

\paragraph{Convolutional neural networks} (CNN) serve as the feature extractor of choice across the included studies: beyond the four studies discussed here, six more use CNNs as FSL technique (see Table~\ref{tab:topics}). Tong and Zhang~\cite{tong2024realTimeLabelFree} combine attention, a customized CNN, an Auto-Encoder and an adversarial discriminator into a self-supervised NIDS. 
Zhang et al.~\cite{zhang2025transferlearning} pre-train and prune a CNN on general traffic. Then, they fine-tune it in a teacher-student twin setup, yielding a lightweight model. Their extensive parameter reporting is a positive example among the included studies. Xu et al.~\cite{xu2025multimodalFusion} fuse two modalities, namely grayscale images of packets handled by a CNN and statistical flow features handled by a transformer. Self-attention fusion worked best. Learned features are then preserved through transfer learning. In addition to OSL, Mirsadeghi et al.~\cite{mirsadeghi2023sdn} utilize sampling and generation (SMOTE \cite{chawla2002smote}, GANs \cite{goodfellow2014gan}) and weighted random forests in the software defined networking (SDN) domain. Regarding OSL, they use a twin CNN, which was outperformed by the weighted random forest regarding precision on some classes.
CNNs bring in mixed results, but overall better than Meta-Learning approaches. Especially \cite{xu2025multimodalFusion} and \cite{yin2026efficient} achieve very good results on popular data sets, making them easily comparable.

\paragraph{Graph Neural Networks} (Graph NN) represent flows or hosts as nodes and their interactions as edges, allowing models to exploit the topology of attacks. Asante and Abass~\cite{asante2026robustMalwareDetection} combine adversarial graph contrastive learning with meta-learning on heterogeneous behavior graphs, achieving the best reported precision and F1-score of the included studies (0.996 on IoT-23). Qiu et al.~\cite{qiu2025disentaglement} trace the inconsistent outcomes of earlier GNNs such as E-GraphSAGE \cite{lo2022egraphsage} back to entangled feature distributions in the graph construction and address this with a five-stage model that disentangles graph representations. Invariants are captured by a projection-based few-shot model. Mao et al.~\cite{mao2025federated} distribute contrastive and classification tasks over federated clients sharing only model weights. Graph approaches perform above average, with $F_1$-score ranging from 0.868 to 0.996. All three evaluate on datasets rarely used by other studies, which limits comparability.

\paragraph{Few-Shot Class-Incremental Learning}  (FSCIL) addresses so-called catastrophic forgetting by adding new classes over time as they appear from few samples. 
Du et al.~\cite{du2024fscilForNIDS} reuse the encoder part of an Auto-Encoder trained on flow-based images within a vision-transformer architecture \cite{dosovitskiy2021viT} and add session-trained projection layers, fusing them. Yin et al.~\cite{yin2026efficient} employ FSL to fine-tune a pre-trained CNN. This FSL is done by constructing a prototypical network using cosine similarity and a CosFace margin \cite{CosFace}, mitigating new attack types through an incremental few-shot process with a specialized loss. Although Yin et al. report no F1-score, their precision of 0.978 is among the best in the included studies.

\paragraph{Auto-Encoders} learn compressed representations of traffic, either for reconstruction"=based pre-training or for generating new samples. 
Hang et al.~\cite{hang2023FlowMAE} pre-train a masked Auto-Encoder on unlabeled traffic and then replace the decoder with a linear classifier for fine-tuning using FSL.
Ayesha et al.~\cite{ayesha2023fs3} chain self-supervised learning with FSL using an Auto-Encoder and contrastive learning and last a kNN classifier. 
Auto-Encoders also appear as learning technique in two other studies \cite{tong2024realTimeLabelFree,du2024fscilForNIDS}. 
This approach, combined with other techniques seems a promising attempt at few-shot learning.

\paragraph{Other Approaches}
Adapting from the field of natural language processing, Ye et al.~\cite{ye2022dirichlet}  propose a text-based workflow based on Dirichlet generative learning, augmenting their few samples with synthetic ones based on semantics.
Martinez-Lopez et al.~\cite{martinez2026multipleSpaces} show that fusing four distance metrics (Chebyshev, Cosine, Euclidean, Wasserstein) outperforms single-metric prototypical few-shot learning. They use these prototypes to generate new datapoints.
Jamshidi et al.~\cite{jamshidi2026thinkFast} base their IoT defense on edge devices. They first construct an anomaly-detection, testing various ML techniques. After a detection, an LLM API call is issued, using the LLM to triage the incident. It is the only one among our included studies, that was built around an LLM.

%% file: 05-discussion-new.tex
\section{Discussion}
\label{sec:discussion}



This review set out to answer three research questions. Regarding \textbf{RQ1} (learning technique), the field is dominated by meta-learning (8 of 21 studies) and convolutional neural networks (10 of 21), which are rarely used in isolation: 17 of 21 studies combine at least two techniques, most frequently meta-learning with CNNs (5 studies). Regarding \textbf{RQ2} (datasets), CIC-IDS2017 and CSE-CIC-IDS2018 have emerged as de facto benchmark datasets (11 and 6 uses), and the reviewed studies evaluate on a median of three datasets each; at the same time, the long tail of single-use datasets means that many studies cannot be compared against any other. Regarding \textbf{RQ3} (evaluation setting), most studies evaluate five or fewer samples per class, with $k=5$ being the most common setting among the studies reporting $k$, and 15 of 21 studies report an $F_1$-score. Reporting is incomplete in a relevant share of studies: five omit $k$, five omit $n$, and most provide neither accessible source code nor full hyperparameters.

\subsection{Interpretation and Implications}

\paragraph{Interpretation.}
Our results support four main observations. First, the most widespread techniques (meta-learning with CNNs or builds on MAML) do not yield the strongest results, whereas graph neural networks and Auto-Encoder-based combinations achieve the highest reported scores. Second, technique adoption appears trend-driven: Auto-Encoders vanish from our set of studies after 2024 despite strong results, and graph neural networks emerge only in 2025--2026. Third, datasets shape reported performance at least as much as techniques on CIC-IDS2017 alone, reported $F_1$-scores range from 0.673 to 0.995 although the convergence on this benchmark family provides a partial basis for comparison.

\paragraph{Implications.}
For practitioners, CNN-based approaches evaluated on the de facto benchmark datasets currently offer the best comparability and hence the most reliable basis for adoption decisions. For researchers, our findings argue for a shared evaluation protocol: fixed $n$-way/$k$-shot settings on common datasets, a defined metric set covering at least accuracy, precision, recall, and $F_1$-score, and mandatory disclosure of code and hyperparameters.

\subsection{Limitations}
Our venue-based quality filter (a journal $h_4$-index of at least 53 according to \cite{ooir}, or a conference ranked B or better in the CORE ranking) was necessary to keep the screening feasible. We acknowledge, however, that citation-based metrics vary considerably between research communities, so relevant work published in newer or lower-ranked venues may have been missed. 

Screening, eligibility decisions, data extraction, and the design of the search queries were carried out by one researcher. Although all decisions are documented through PRISMA-like reporting, which makes them traceable, a second reviewer would have reduced the subjectivity.

Finally, our synthesis is limited by the primary studies themselves: missing parameters and unavailable source code prevented us from verifying reported numbers, so our performance comparison relies on self-reported results obtained under heterogeneous settings.

\subsection{Future Work}
Two directions for future work follow directly from our findings. 
First, a benchmark study that re-implements representative approaches from each technique family and evaluates them under a unified protocol with fixed $n$ and $k$, common datasets, identical preprocessing, and a full metric set. This would turn the indications observed here into verifiable rankings. Publishing such a benchmark as a community artifact, together with a reporting checklist for few-shot NIDS studies, would directly address the reproducibility gaps identified above.
Second, generalization across datasets remains essentially untested: none of our included studies trains and evaluates on disjoint datasets, so the robustness of these approaches to unseen network environments is unknown.

%% file: 06-conclusion.tex
\section{Conclusion}
\label{sec:conclusion}

Few-shot learning addresses a central obstacle to learning-based intrusion detection: the scarcity of labeled attack data. In this work, we systematically reviewed recent few-shot learning approaches for NIDS using records retrieved from the ACM Digital Library, IEEE Xplore, and Scopus. Starting from 1,358 records published between 2022 and August 2026, we retained 21 peer-reviewed studies and analyzed their learning techniques, datasets, and evaluation settings.

Our analysis shows a field dominated by convolutional neural networks and meta-learning, typically combined with other techniques. The datasets CIC-IDS2017 and CSE-CIC-IDS2018 have emerged as de facto benchmarks and provide a partial basis for comparison. However, heterogeneous evaluation settings, incomplete parameter reporting, and largely unavailable source code still limit direct comparison and independent verification of reported results.

For few-shot NIDS to mature into deployable defenses, the community needs shared evaluation protocols and stronger reproducibility practices. We therefore advocate fixed n-way/k-shot settings on common datasets, consistent reporting of evaluation parameters and metrics, and the disclosure of source code and hyperparameters. A unified, openly reproducible benchmark, ideally complemented by cross-dataset evaluation, would be the most valuable next step for assessing both performance and generalization.




\paragraph{Data availability statement} The data and scripts required to reproduce our results are available at \url{https://speicherwolke.uni-leipzig.de/index.php/s/AFTL6xKRPaA3BfW}. 

%% file: 07-acknowledgements.tex
\section{Acknowledgment}
\label{sec:acknowledgment}
The authors acknowledge the financial support by the Federal Ministry of Research, Technology and Space of Germany and by Sächsische Staatsministerium für Wissenschaft, Kultur und Tourismus in the programme Center of Excellence for AI-research „Center for Scalable Data Analytics and Artificial Intelligence Dresden/Leipzig“, project identification number: ScaDS.AI

%% file: 08-appendix.tex
\section*{Appendix}
\label{sec:appendix}


\begingroup
\footnotesize
\setlength{\tabcolsep}{5pt}
\begin{longtable}{@{}L{1.8cm}L{9.5cm}@{}}

\caption{Database-specific search queries.}
\label{tab:queries}\\
\toprule
\textbf{Source} & \textbf{Query} \\
\midrule
\endfirsthead

\multicolumn{2}{@{}l}{\tablename~\thetable\ \emph{(continued)}}\\[2pt]
\toprule
\textbf{Source} & \textbf{Query} \\
\midrule
\endhead

\midrule
\multicolumn{2}{r@{}}{\emph{Continued on next page}}\\
\endfoot

\bottomrule
\endlastfoot

ACM &
\texttt{[[All: ids] OR [All: intrusion detection] OR [All: intrusion
prevention] OR [All: idps]] AND [[All: learning] OR [All: few shot
learning] OR [All: fsl] OR [All: one-shot learning] OR [All: one shot
learning] OR [All: osl]] AND [E-Publication Date: (01/01/2022 TO
08/31/2026)]} \\
\addlinespace
IEEE Xplore &
\texttt{(("All Metadata":intrusion detection) OR ("All Metadata":ids)
OR ("All Metadata":intrusion prevention) OR ("All Metadata":idps)) AND
(("All Metadata":learning) OR ("All Metadata":few shot learning) OR
("All Metadata":one shot learning) OR ("All Metadata":fsl) OR
("All Metadata":one-shot learning) OR ("All Metadata":osl))} \\
\addlinespace
Scopus &
\texttt{(TITLE-ABS-KEY("intrusion detection") OR TITLE-ABS-KEY("ids")
OR TITLE-ABS-KEY("intrusion prevention") OR TITLE-ABS-KEY("idps")) AND
(TITLE-ABS-KEY("few shot learning") OR TITLE-ABS-KEY("one shot
learning") OR TITLE-ABS-KEY("fsl") OR TITLE-ABS-KEY("osl") OR
TITLE-ABS-KEY("learning") OR TITLE-ABS-KEY("one-shot-learning")) AND
LIMIT-TO(LANGUAGE, "English") AND PUBYEAR > 2022 AND PUBYEAR < 2026} \\

\end{longtable}
\endgroup

\newlength{\yearcol}\setlength{\yearcol}{0.8cm}
\newlength{\papercol}\setlength{\papercol}{3.7cm}
\newlength{\ncol}\setlength{\ncol}{1.0cm}
\newlength{\kcol}\setlength{\kcol}{1.5cm}
\newlength{\metriccol}\setlength{\metriccol}{1.35cm}
\newlength{\datasetcol}

\begin{landscape}
\begingroup
\footnotesize
\setlength{\datasetcol}{\dimexpr\linewidth-\yearcol-\papercol-\ncol-\kcol-4\metriccol-18\tabcolsep\relax}
\begin{longtable}{P{\yearcol} L{\papercol} L{\datasetcol} P{\ncol} P{\kcol} *{4}{P{\metriccol}}}
\caption{Datasets, parameters and results. Results averaged over all datasets of a study are marked with * and balanced accuracy with **. Macro-averaged scores are marked with an indicative (M).}
\label{tab:results}\label{tab:datasets}\\
\toprule
Year & Author & Dataset & n & k & Accuracy & Precision & Recall & F1-Score\\
\midrule
\endfirsthead
Year & Author & Dataset & n & k & Accuracy & Precision & Recall & F1-Score\\
\midrule
\endhead
\midrule
\multicolumn{9}{r}{\itshape Continued on next page}\\
\endfoot
\bottomrule
\endlastfoot
2022 & \multirow{2}{=}{Ye et al.~\cite{ye2022dirichlet}} & \text{\color{jungleteal}USTC-TFC2016} & \multirow{2}{*}{18} & \multirow{2}{*}{1,10} &  &  &  & 0.951(M) \\ \nopagebreak
 &  & \text{\color{blushedbrick}CIC-IDS2017} &  &  &  &  &  & 0.962(M) \\
\midrule
2023 & \multirow{3}{=}{Ayesha et al.~\cite{ayesha2023fs3}} & WuSTl-EHMS & 2 & 5,10 &  & 0.979 & 0.981 & 0.980 \\ \nopagebreak
 &  & WuSTl-IIoT & 5  & 5,10  &  & 0.889 & 0.680 & 0.701 \\ \nopagebreak
 &  & CIC-BoT-IoT & 5 & 5,10 &  & 0.619 & 0.603 & 0.596 \\
\cmidrule(lr){2-9}
 & \multirow{6}{=}{Hang et al.~\cite{hang2023FlowMAE}} & \text{\color{sandybrown}CSE-CIC-IDS2018} & & & 0.999 & 0.999 & 0.999 & 0.999 \\ \nopagebreak
 &  & \text{\color{jungleteal}USTC-TFC2016} &  &  & 0.998 & 0.998 & 0.998 & 0.998 \\ \nopagebreak
 &  & ISCX-VPN-2016 (App) &  &  & 0.998 & 0.999 & 0.998 & 0.999 \\ \nopagebreak
 &  & ISCX-VPN-2016 (Service) &  &  & 0.991 & 0.992 & 0.991 & 0.991 \\ \nopagebreak
 &  & ISCX-Tor-2016 &  &  & 0.994 & 0.994 & 0.994 & 0.994 \\ \nopagebreak
 &  & Cross-platform &  &  & 0.992 & 0.992 & 0.992 & 0.992 \\
\cmidrule(lr){2-9}
 & Lu et al.~\cite{lu2023MAML_IDS} & FSIDS-IoT & 5 & 1,5,10 & 0.896 &  &  &  \\
\cmidrule(lr){2-9}
 & Mirsadeghi et al.~\cite{mirsadeghi2023sdn} & InSDN & 8 & 1 &  &  &  & 0.700(M) \\
\cmidrule(lr){2-9}
 & \multirow{3}{=}{Miao et al.~\cite{miao2023spn}} & ISCX-IDS2012 &  & \multirow{3}{*}{1,3,5,10} & 0.983 &  &  &  \\ \nopagebreak
 &  & \text{\color{blushedbrick}CIC-IDS2017} &  &  & 0.996 & 0.924(M) & 0.937(M) & 0.924(M) \\ \nopagebreak
 &  & \text{\color{jungleteal}USTC-TFC2016} &  &  & 0.991 & 0.965(M) & 0.966(M) & 0.965(M) \\
\cmidrule(lr){2-9}
 & Sun et al.~\cite{sun2023capsuleNetwork} & \text{\color{sandybrown}CSE-CIC-IDS2018} & 8 &  & 0.952 &  &  & 0.988(M) \\
\midrule
2024 & \multirow{3}{=}{Tong and Zhang~\cite{tong2024realTimeLabelFree}} & NSL-KDD & 5 &  & 0.923 & 0.940 & 0.887 & 0.913 \\ \nopagebreak
 &  & \text{\color{blushedbrick}CIC-IDS2017} & 7 &  & 0.997 & 0.993 & 0.997 & 0.995 \\ \nopagebreak
 &  & \text{\color{sandybrown}CSE-CIC-IDS2018} & 7 &  & 0.996 & 0.997 & 0.997 & 0.997 \\
\cmidrule(lr){2-9}
 & \multirow{2}{=}{Du et al.~\cite{du2024fscilForNIDS}} & \text{\color{blushedbrick}CIC-IDS2017} & 12 & \multirow{2}{*}{5} & 0.931 & 0.931 & 0.931 & 0.929 \\ \nopagebreak
 &  & \text{\color{sandybrown}CSE-CIC-IDS2018} & 14 &  & 0.901 & 0.909 & 0.901 & 0.897 \\
\midrule
2025 & \multirow{3}{=}{Mao et al.~\cite{mao2025federated}} & NF-BoT-IoT-v2 & 5 & & 0.984 & 0.980 & 0.799 & 0.868 \\ \nopagebreak
 &  & NF-ToN-IoT-v2 & 10 &  & 0.943 & 0.742 & 0.724 & 0.733 \\ \nopagebreak
 &  & NF-CSE-CIC-IDS2018-v2 & 7 &  & 0.995 & 0.844 & 0.790 & 0.813 \\
\cmidrule(lr){2-9}
 & \multirow{3}{=}{Zhang et al.~\cite{zhang2025transferlearning}} & Edge-IIoT & \multirow{3}{*}{5} & \multirow{3}{*}{1,5,10,20} &  &  &  &  \\ \nopagebreak
 &  & CIC-BoT-IoT &  &  & 0.922* &  &  &  \\ \nopagebreak
 &  & \text{\color{jungleteal}USTC-TFC2016} &  &  &  &  &  &  \\
\cmidrule(lr){2-9}
 & \multirow{2}{=}{Xu et al.~\cite{xu2025multimodalFusion}} & \text{\color{blushedbrick}CIC-IDS2017} & \multirow{2}{*}{2,3,4} & \multirow{2}{*}{5,10,15} & 0.934 &  &  &  \\ \nopagebreak
 &  & \text{\color{sandybrown}CSE-CIC-IDS2018} &  &  & 0.985 &  &  &  \\
\cmidrule(lr){2-9}
 & \multirow{3}{=}{Xu et al.~\cite{xu2025mutalCentralizedLearning}} & ISCX-IDS2012 & \multirow{3}{*}{2,4,5} & \multirow{3}{*}{5,10} & 0.882 &  &  & 0.883 \\ \nopagebreak
 &  & \text{\color{blushedbrick}CIC-IDS2017} &  &  & 0.922 &  &  & 0.922 \\ \nopagebreak
 &  & \text{\color{sandybrown}CSE-CIC-IDS2018} &  &  & 0.887 &  &  & 0.884 \\
\cmidrule(lr){2-9}
 & \multirow{5}{=}{Qiu et al.~\cite{qiu2025disentaglement}} & CIC-ToN-IoT &  & \multirow{5}{*}{5} &  &  &  & 0.961 \\ \nopagebreak
 &  & CIC-BoT-IoT &  &  &  &  &  & 0.982 \\ \nopagebreak
 &  & Edge-IIoT &  &  &  &  &  & 0.968 \\ \nopagebreak
 &  & NF-UNSW-NB15-v2 &  &  &  &  &  & 0.954 \\ \nopagebreak
 &  & NF-CSE-CIC-IDS2018-v2 &  &  &  &  &  & 0.963 \\
\midrule
2026 & \multirow{4}{=}{Martinez-Lopez et al.~\cite{martinez2026multipleSpaces}} & \text{\color{blushedbrick}CIC-IDS2017} &  & \multirow{4}{*}{2} & 0.880** &  &  & 0.673 \\ \nopagebreak
 &  & CIC-EVSE2024 Network &  &  & 0.820** &  &  & 0.850 \\ \nopagebreak
 &  & CIC-EVSE2024 PowerB &  &  & 0.738** &  &  & 0.552 \\ \nopagebreak
 &  & CIC-IoV2024 &  &  & 0.887** &  &  & 0.754 \\
\cmidrule(lr){2-9}
 & \multirow{3}{=}{Wu et al.~\cite{wu2026negativeLearning}} & \text{\color{blushedbrick}CIC-IDS2017} & \multirow{3}{*}{5} & \multirow{3}{*}{5,10} & 0.743 &  &  &  \\ \nopagebreak
 &  & Edge-IIoT &  &  & 0.962 &  &  &  \\ \nopagebreak
 &  & IoT-23 &  &  & 0.840 &  &  &  \\
\cmidrule(lr){2-9}
 & \multirow{2}{=}{Yin et al.~\cite{yin2026efficient}} & \text{\color{blushedbrick}CIC-IDS2017} & \multirow{2}{*}{5} & \multirow{2}{*}{5} & 0.998 & 0.978 & 0.981 &  \\ \nopagebreak
 &  & \text{\color{jungleteal}USTC-TFC2016} &  &  & 0.983 & 0.918 & 0.903 &  \\
\cmidrule(lr){2-9}
 & Jamshidi et al.~\cite{jamshidi2026thinkFast} & \text{\color{blushedbrick}CIC-IDS2017} & 15 &  & 0.989 &  &  &  \\
\cmidrule(lr){2-9}
 & \multirow{2}{=}{Zhang et al.~\cite{zhang2026maml-samIoT}} & FSIDS-IoT & \multirow{2}{*}{5} & \multirow{2}{*}{1,5,10} & 0.915 & 0.914 & 0.915 & 0.915 \\ \nopagebreak
 &  & FSIDS-IoT-v2 &  &  & 0.905 & 0.905 & 0.905 & 0.905 \\
\cmidrule(lr){2-9}
 & Asante and Abass~\cite{asante2026robustMalwareDetection} & IoT-23 & 5 & 1,5 & 0.997 & 0.996 & 0.997 & 0.996 \\
\cmidrule(lr){2-9}
 & \multirow{3}{=}{Lu et al.~\cite{lu2026dmaml}} & \text{\color{blushedbrick}CIC-IDS2017} & \multirow{3}{*}{2,5} & \multirow{3}{*}{5} & 0.902 & 0.892 & 0.875 & 0.883 \\ \nopagebreak
 &  & \text{\color{oliv}UNSW-NB15} &  &  & 0.967 & 0.960 & 0.954 & 0.957 \\ \nopagebreak
 &  & CIC-ToN\_IoT &  &  & 0.969 & 0.963 & 0.958 & 0.960 \\
\end{longtable}
\endgroup
\end{landscape}